\documentclass{cccg18}
\usepackage{graphicx,amssymb,amsmath, mathtools}
\usepackage{url}
\usepackage{subfig}
\usepackage[title]{appendix}

\newcommand{\reals}{{\rm I\!\hspace{-0.025em} R}}

\title{Dihedral Rigidity and Deformation}

\author{Nina Amenta\thanks{Department of Computer Science,
        University of California, Davis, {\tt amenta@cs.ucdavis.edu}}
        \and
        Carlos Rojas\thanks{Department of Computer Science,
        University of California, Davis, {\tt  crojas@ucdavis.edu}}}

\index{Amenta, Nina}
\index{Rojas, Carlos}

\begin{document}
\thispagestyle{empty}
\maketitle

\begin{abstract}
We consider defining the embedding of a triangle mesh into
$\reals^3$, up to translation, rotation, and scale, by its vector of dihedral angles.
Theoretically,
we show that locally, almost everywhere, the map from realizable vectors of dihedrals to
mesh embeddings is one-to-one.
We experiment with a heuristic
method for mapping straight-line interpolations in dihedral
space to interpolations between mesh embeddings and produce smooth and intuitively appealing morphs between three-dimensional shapes.
\end{abstract}

\section{Introduction}

Frequently, a polygon mesh is represented by its mesh combinatorics and a vector of
3D vertex coordinates, specifying the immersion of the mesh into $\reals^3$.
A polygon mesh is \textit{rigid} when the only motions of the vertex coordinates
for which the faces are not deformed in any way are the
rigid motions (rotation and translation).
Non-rigid polyhedra do exist \cite{bricard,connelly,connelly2},
although they are rare; a non-rigid polyhedron has some \textit{flexing}
motion in which the faces move but do not deform.
That is, the dihedral angles between
faces change continuously while
the faces themselves remain rigid.
In 1974 Herman Gluck~\cite{gluck} proved
\begin{theorem}[Gluck]
A generic immersion of any mesh topology homeomorphic to the sphere is rigid.
\end{theorem}
By \textit{generic} we mean all vectors of vertex coordinates, except some
``degenerate" subset of measure zero.
So, for example, if you
construct the
edge-skeleton of triangulated computer graphics model,
with stick edges held
together at flexible joints, it
almost certainly would be rigid.

Mesh deformations, in which, typically, both edge lengths and dihedrals change, is an important topic in computer graphics, computer vision and scientific shape analysis.  A deformation defines a path in the space of discrete metrics - a continuous change in the vector of edge lengths. This has been quite well studied, especially for the subset of discrete conformal transformations, eg. \cite{sorkine2007rigid, baek2015isometric, zhang2015fast}.
But discrete metrics
do not correspond to shape in any precise sense:
rigidity theory tells us that a vector of edge lengths typically
has multiple discrete realizations as a rigid mesh, and that it might even correspond to a
flexible polyhedron.

We are interested in the other possibility: characterizing a deformation by the change in its vector of dihedral angles.
The first mathematical question
one might ask is whether there are
motions in which all the dihedral angles stay the same, but the edge lengths change.
Indeed, this is trivially possible; consider a cube deforming into an arbitrary box.
But notice that during such a deformation the inner face angles (the plane angles) remain unchanged.
So next we ask if there are deformations in which the dihedral angles remain fixed,
but the inner angles change;
we call this a \textit{dihedral flex}.
We say that a polyhedron which does \textit{not} allow a dihedral flex is \textit{dihedral-rigid}.
It is not known if dihedral-flexible polyhedra exist.
Here, we prove the following analog of Gluck's theorem:
\begin{theorem}
\label{thm:dihedral}
A generic immersion of any triangle
mesh homeomorphic to the sphere is dihedral-rigid.
\end{theorem}
While interesting as a result in rigidity theory,
of course this is only a small step towards a theoretical justification of the idea of representing mesh embeddings by their dihedral vectors.
We also give some experimental evidence that the dihedral representation is useful and natural.
We compute a morphs between two embeddings
by heuristically mapping the straight line segment connecting their dihedral vectors in (Euclidean) dihedral-space onto a path in the space of embeddings.
We find smooth paths
connecting very different shapes, and observe that the resulting morphs seem quite natural.

\section{Related work}

In 1968, Stoker \cite{stoker} conjectured that a convex polyhedron is uniquely defined by 
its combinatorics and dihedral angles (and thus that it is dihedral-rigid).
This would be the dihedral version of Cauchy's theorem on the rigidity of convex 
polyhedra \cite{cauchy}. 
A fairly simple proof of Stoker's conjecture for triangulated convex polyhedra was given by 
Pogorelov \cite{pogo}; we draw on his
work as well as that of Gluck. 
Only recently was a complete proof
of Stoker's conjecture
provided by Mazzeo and Montcouquiol \cite{mnm}, 
using much more sophisticated techniques and applying to the interesting case of ideal
hyperbolic polyhedra as well. 

In computer graphics, there is 
an ongoing interest in 
constructing shape spaces in which geodesic paths correspond to physically natural-looking morphs, which can be
used for applications such as 
morphing, shape exploration, deformation and deformation transfer.
These spaces tend to be curved and
difficult to deal with, eg.
\cite{kilian2007geometric}. 
A recent series of papers \cite{heeren2014exploring,heeren2016splines,zhang2015shell}
explores the curved shape space implied by the elastic model of deformation.
They
prove that it forms a Riemannian manifold, and
produce shape averages, principal components and splines in this ``shell space".
Each of these operations proves to be challenging,
both mathematically and computationally.

There is a practically
successful line of work 
\cite{baek2015isometric,kircher2008free,winkler2010multi}
on interpolating mesh embeddings by interpolating
both their dihedral angles and their edge lengths, and then doing some sort of least-squares reconstruction to 
produce an interpolating mesh. 
These methods cannot 
realize both the 
dihedrals and the edge lengths exactly - there are roughly $6n$ parameters and 
$3n$ degrees of freedom in the embedding, where $n$ is the number of 
mesh vertices - but they are fairly simple and 
they provide very nice-looking results.

The space of dihedral angles was proposed recently as a representation for deformation by Paill´e et al.~\cite{paille2015dihedral}, albeit for a tetrahedralized volume. 
Here again, we find that the number of dihedrals in a tetrahedralization is much larger than the dimension of the space of realizable meshes.  
Finally, \cite{isenburg2001connectivity} showed that ignoring edge length and just using connectivity to reconstruct shapes is surprisingly successful.

\section{Infinitesimal rigidity}

One's first instinct when considering the possibility of a dihedral flex is to consider the vertex positions
$p_i$ as functions $p_i(t)$ of some parameter $t$, and consider the derivatives of the inner angles
$\beta_{j,i,j+1}$ and dihedrals $\alpha_{i,j}$  with respect to $t$.
At any point along any traditional (edge length) flex, the length derivatives
$l'_{i,j} = 0$ at every edge, while at least some of the $\alpha'_{i,j}$ are non-zero.
Similarly along any dihedral flex (if such a thing exists!) we expect to find an infinitesimal motion
such that all $\alpha'_{i,j} = 0$,
while there are inner angles for which the derivatives
$\beta'_{j,i,j+1}$ are non-zero.
We call a polyhedron which admits such a motion \textit{dihedral infinitesimally non-rigid}.

A polyhedron which is dihedral non-rigid must be dihedral infinitesimally non-rigid.
It may well be possible, however, for a polyhedron to be dihedral infinitesimally non-rigid while being rigid; there are many polyhedra which are (length)
infinitesimally non-rigid, but actually
rigid.
Following Gluck, we prove that a generic
immersion of a mesh forms a polyhedron which is
dihedral infinitesimally rigid, and hence dihedral rigid.

\section{Dihedral infinitesimal rigidity as a matrix equation}

There is a very nice relationship between the derivatives of the dihedral angles $\alpha'$ and the derivatives of the
triangle inner angles $\beta'$, which
Gluck used in his theorem on length rigidity.
We have, going around the one-ring of any vertex $p_i$,
$$
\sum_{j} \alpha'_{ij} \vec{e}_{ij} + \sum_j \beta'_{j,i,j+1} \vec{n}_{j,i,j+1}  = \mathbf{0}
$$
where $\vec{n}_{j,i,j+1}$ is the normal to triangle $t_{j,i,j+1}$,
and $\vec{e}_{ij} = (p_i - p_j)/{|| p_i - p_j ||}$ is the unit vector in the direction of edge $e_{ij}$.
This equation expresses the fact that the instantaneous angular velocities in the one-ring
have to change in a coordinated fashion for the one-ring to continue to ``hold together".
Their derivation appears in Appendix~\ref{sec:vertEqn}.
Since the edge and normal vectors have three coordinates each, we have three equations
at each vertex, for a total of $3V$.
Let's call these the vertex equations.

Gluck considered the case in which we assume that the change in edge lengths,
and hence the inner angle
derivatives $\beta'$, are all zero, so that the length infinitesimally non-rigid configurations
were those with
$$
\sum_{j} \alpha'_{ij} \vec{e}_{ij}   = \mathbf{0}
$$
This system has $3V$ equations in $3V-6$ variables.

We make the opposite assumption, that the dihedral angles $\alpha$ remain unchanged, so
we are interested in non-zero solutions to
$$
\sum_{j} \beta'_{j,i,j+1} \vec{n}_{j,i,j+1}  = \mathbf{0}
$$
In our case we have $3V$ equations in the $6V-12$ variables $\beta'$.
There are additional constraints on the $\beta'$ variables which determine the validity of
the mesh.

One is that the sum of the inner angles of any triangle add up to $\pi$.
Taking the derivative of this condition is gives us
$$
\beta'_i + \beta'_j + \beta'_{j+1} = 0
$$
We call these the face equations.

Finally, the Law of Sines implies the following differential cotangent formula for the triangles
around a given one-ring
$$
\sum_j \cot \beta_{i,j,j+1} \beta'_{i,j,j+1} - \cot \beta_{i,j+1,j}  \beta'_{i,j+1,j} = 0
$$
The derivation of this equation appears in Appendix~\ref{sec:cotVert}.
We call these the cotangent equations.
Together, the vertex equations, face equations and cotangent equations form a system
with $3V+2V-4+V = 6V-4$ equations in $6V-12$ variables.
$$
M \beta' = \mathbf{0}
$$
A mesh is dihedral infinitesimally non-rigid if this system $M$ has some non-zero solution
for the $\beta$s, that is, if there is an infinitesimal motion of the mesh that leaves the
dihedrals fixed but allows the inner angles to flex somehow, while maintaining a valid mesh.

\section{Condition for a solution}

Following Gluck, we observe that
there is a non-zero solution for $\beta$ if and only if the coefficient matrix
$M$ has rank less than $6V-12$.
And for this to be true, it must be the case that every $6V-12 \times 6V-12$ sub-matrix of $M$
has zero determinant.
We can write this condition on the coefficient matrix itself as a system of
$\binom{6V-4}{8}$ polynomials in the matrix elements;
call this system $F$.
In Gluck's proof, he dealt with a matrix whose coefficients were themselves
polynomials in the vertex coordinates of the
mesh, and this allowed him to argue that the resulting variety formed a set of measure zero.

In our case, the coefficients are the face normals, ones, and the cotangents of the inner angles.
These are not all polynomials in the vertex coordinates.
To get around this, we treat the normals and cotangents as variables themselves; for
notational clarity, let's write $c_{j,i,j+1} = \cot \beta_{j,i,j+1}$.
The $c$ and $n$ variables are not independent of each other.
The normals must all have length one; for $n_{j,i,j+1} = (n_x,n_y,n_z)$, we have
\begin{eqnarray}
\label{eqn:norm}
n_x^2 + n_y^2 + n_z^2 = 1
\end{eqnarray}
In addition, the normal and cotangent variables are conveniently related to each other,
and to the vertex coefficients, by the following formula.

\begin{eqnarray}
\label{eqn:cot}
\begin{aligned}
&\left[ (p_i - p_j) \times (p_i - p_{j+1}) \right]  c_{j,i,j+1} = \\
&\left[ (p_i - p_j) \cdot (p_i - p_{j+1}) \right]   n_{j,i,j+1}
\end{aligned}
\end{eqnarray}

This formula relates the cotangent to the
scaling of the cross-product to form
the triangle normal; an
(easy) derivation appears in Appendix~\ref{sec:cot}.
Note that since the cross-product and dot-product are both polynomial functions, this is
a polynomial as well.

In order for a mesh configuration to be be dihedral infinitesimally non-rigid, we need
Equations \ref{eqn:norm} and \ref{eqn:cot}
to be true for every angle, as well as for all of the sub-determinants of $M$ to
be zero.
These conditions are all polynomial, and they define a variety (the intersection of their zero-sets) in
the space of the $p,n,c$ variables.

An arbitrary assignment of values to
$p,n,c$ does not correspond to an
immersion of the
mesh;
the $p$-variables are all free, but
the $n$ and $c$ will not obey Equations \ref{eqn:norm} and \ref{eqn:cot}.
Given a choice of $p$ variables, the
$n$ and $c$ variables of that embedding uniquely
satisfy \ref{eqn:norm} and \ref{eqn:cot}
(the normal is indeed the cross-product,
scaled as required).
So there is unique lifting of the Euclidean space defined by the vertex coordinate space $p$
into $(p,n,c)$-space.
Let $\tilde{P}$ be this lifting of the
the vertex coordinate space, which is Euclidean.
The space $\tilde{P}$ is similarly simply
connected and $3n$-dimensional.

If we have a connected component of an algebraic variety and we add an
additional polynomial constraint to the system, either the
the whole component satisfies the new equation,
or the dimension of the new variety is reduced
by the intersection with the new equation.
Thus, if there is any point of $\tilde{P}$ which does \textit{not} also satisfy the
system $F$ saying that all of the sub-determinants have to be zero, the
set of common zeros (the space of dihedral infinitesimally non-rigid polyhedra) has smaller
dimension than $\tilde{P}$, and forms a subset of measure zero.

So all we need to do to show that the dihedral infinitesimally non-rigid polyhedra form
a set of measure zero is to display some point in $\tilde{P}$ which is \textit{not} in
$F$; that is, a dihedral infinitesimally rigid polytope.
As it happens, we can do this for every mesh topology; the results of
Pogorelov~\cite{pogo} and Mazzeo and Montcouquiol~\cite{mnm} show that
every convex polyhedron is dihedral infinitesimally rigid, and we know that every
mesh topology can be realized as a convex polyhedron (this is Steinitz' theorem).

This proves Theorem~\ref{thm:dihedral}.

\section{Experiments with dihedral parameterization}
\label{sec:dihedral_exp}

\begin{figure*}[tbh]
\centering
\includegraphics[width=0.9\textwidth]{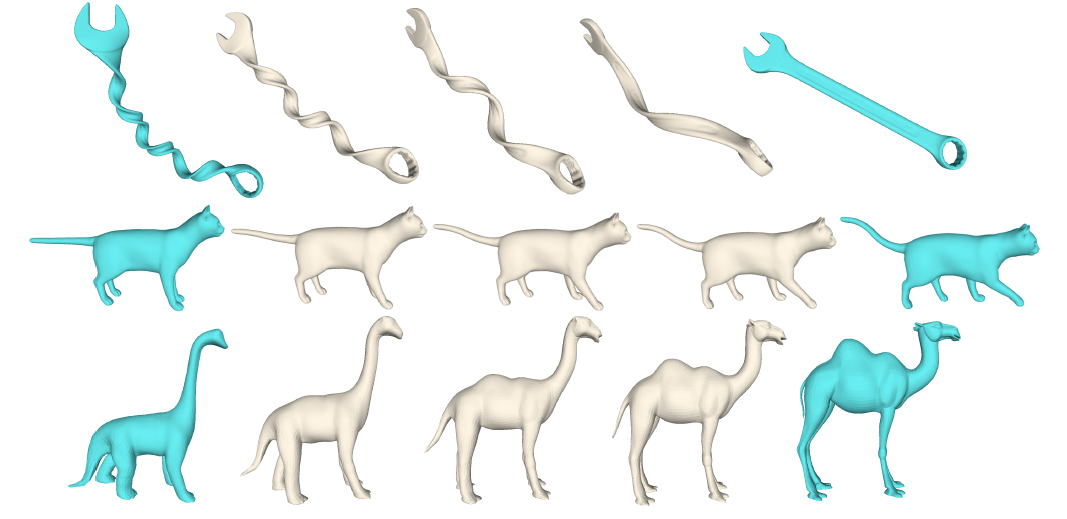}
\caption{Three examples of morphs between different shapes.  The twisting of the wrench is handeled nicely by considering dihedrals.  The poses of the cat are interpolated naturally, without distortion in the intermediate shapes.  The dinosaur and the camel 
have the same mesh topology, but are very different embeddings.}
\label{fig:dino}
\end{figure*}

In this section we experiment with treating the dihedral vector for a given mesh topology as a Euclidean shape space. First, we consider interpolating between different embeddings of the same mesh by interpolating 
their dihedral angles.
We find that this produces smooth morphs between quite different shapes. For instance, in Figure~\ref{fig:dino}, we get a smooth morph from a dinosaur to a camel. 

We morph between the two input embeddings by connecting their two dihedral vectors with a straight line segment in dihedral space, 
and reconstructing a series of embeddings 
corresponding to uniformly-spaced points along the segment.
Since the dihedrals are scale-invariant, 
we normalize the scale of the 
embedding as
well as its rotation and translation. 
This means that the space of possible embeddings had dimension $3n-7$, where 
$n$ is the number of mesh vertices
($3n$ possible vertex coordinates, normalized
for the seven-dimensional transformation space). 
A mesh homeomorphic to the sphere has 
$3n-6$ dihedral angles, however, so 
we do not expect the intermediate dihedral 
vectors to exactly correspond to embeddings.
Instead, we use a 
least-squares algorithm to compute embeddings that lie as close as possible to the 
dihedral-space line segment connecting the input shapes.

At each interpolated point, 
we use an
optimization algorithm to find a 
mesh embedding that 
comes as close as possible to realizing
the desired dihedrals.  
The optimization algorithm consists of 
an initialization step and a
refinement step. 
The initialization fits an embedding 
to both interpolated dihedrals and interpolated edge lengths. 
Then 
a refinement step
alternates between computing a set of normal vectors $\vec{n}_k$ which realize the given dihedrals, 
and a set of mesh vertex 
positions $p_i$ which 
realize the $\vec{n}_k$ as
well as possible. 
The initialization, and each iteration 
of the refinement process, consists of
a linear least-squares solve.

The initialization step is 
similar to the 
mesh interpolation algorithms 
of Baek et al.~\cite{baek2015isometric} and 
Kircher and Garland~\cite{kircher2008free}. We first reconstruct
the one-ring of each vertex, given the interpolated edge lengths and dihedral, and then we combine the one-rings, 
using least-squares, to produce a set of vertex positions.  We describe 
this in 
more detail in Appendix~\ref{init}. 

The refinement phase is more novel.
We define an energy function $E$ for a mesh, which considers both the normal vectors $\vec{n}_k$ and the vertices $p_i$.

\begin{equation}
\begin{aligned}
\label{eqn:dihedral_energy}
E = \alpha \sum_{\mathclap{\mbox{adjacent triangles } k,l}} || M_{kl} \vec{n}_k - \vec{n}_l ||_F^2 + \\
\beta \sum_{\mathclap{\substack{{\mbox{edge } i,j }\\ \mbox{triangle } k}}}  || L_{ijk} (p_i - p_j) - \vec{n}_k||^2
\end{aligned}
\end{equation}

$F$ indicates the Frobenious norm. 
Here $\vec{n}_k$ is the normal of triangle $k$ and $\vec{n}_l$ is the normal of triangle $l$, adjacent across edge $i,j$. 
The matrix $M_{kl}$ is a rotation by exactly the desired dihedral angle $\delta_{ij}$ between $\vec{n}_k$ and $\vec{n}_l$, with the axis of rotation $\vec{e}_{ij} = (p_i - p_j)/{|| p_i - p_j ||}$.
Thus the first energy term measures how well the normals achieve the dihedral angles at every edge.
The second term measures how well the normals and vertices agree with each other. 
The matrix $L_{i,j,k}$ takes edge $i,j$ into the normal of one of its adjacent 
triangles $k$. 
It is the product of a rotation by 
$\pi/2$, along with a scaling to 
normalize the length. 
The weights are $\alpha = 0.6$ and 
$\beta=0.4$.

At each step, we recompute $M$ and $L$
from the current mesh, solve for new $\vec{n}_k$ while keeping the $p_i$ fixed, and finally solve for new values of 
$p_i$. 

\begin{figure}[tbh]
\centering
\includegraphics[width=0.5\textwidth]{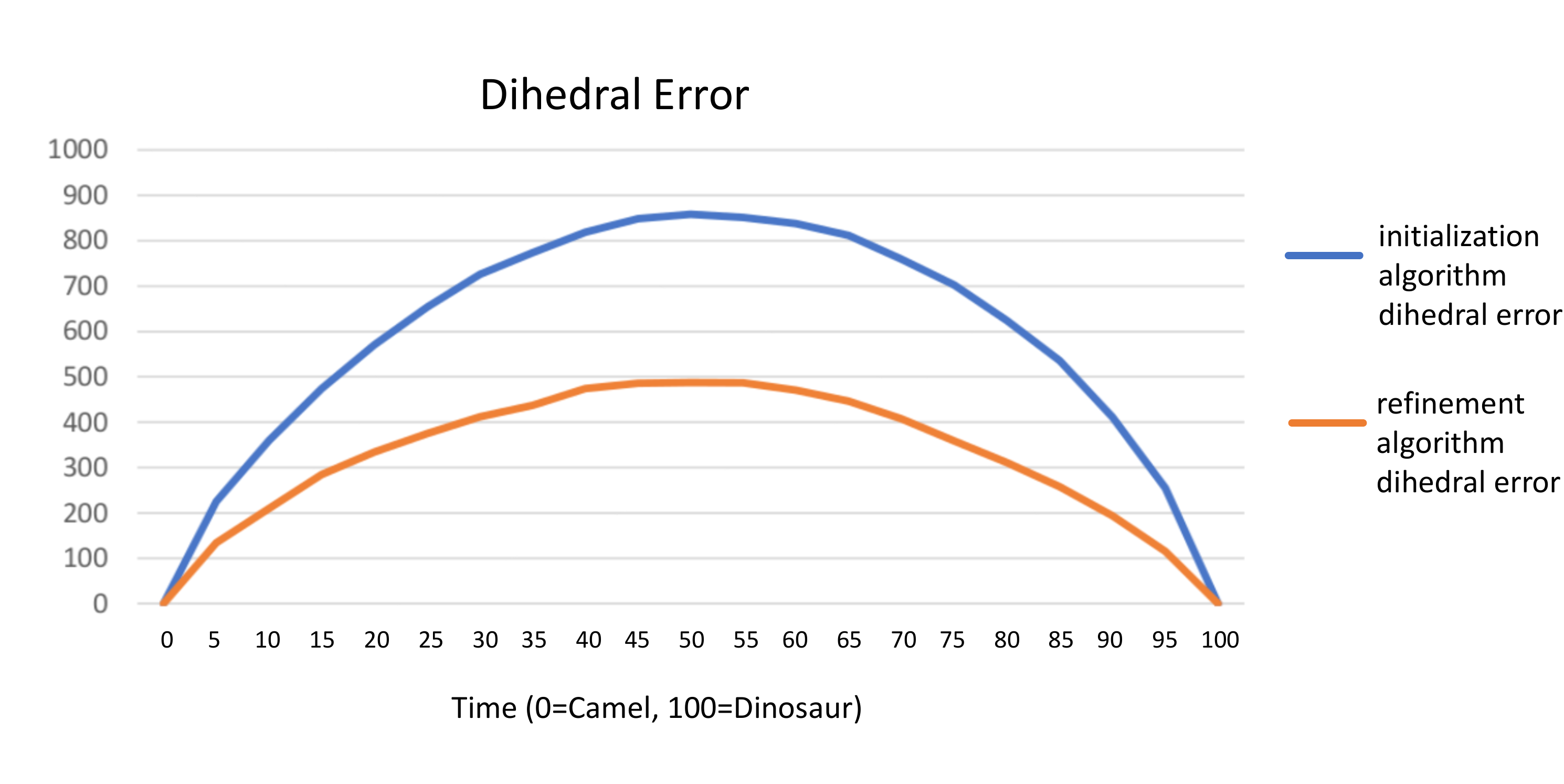}
\caption{Dihedral error reduction 
due to the refinement step, for the camel-to-dinosaur morph.}
\label{errorChart}
\end{figure}

We see that this algorithm succeeds in reducing the 
dihedral error at of the interpolations
by about half.  
We define the 
dihedral error simply 
as the 
Euclidean difference between the desired interpolated dihedral vector and 
the actual dihedrals achieved by our embedding; an example
appears in Figure~\ref{errorChart}.  
As noted above, we do 
not expect to be able to achieve 
a dihedral error of zero.  

  Videos of these smooth morphs can be seen at \url{https://vimeo.com/270302684}.

In earlier experiments \cite{rojas2014}, we found that attempting to optimize the embedding towards the interpolated edge lengths, rather than the interpolated dihedrals, produced morphs with discontinuities and glitches.  We believe that this is because there are many possible embeddings realizing a given set of edge lengths, while we suspect that at most one embedding per vector of dihedrals. 

\section{Shape analysis}

We also considered using the space
of dihedral angles as a method for
shape analysis. This idea is 
appealing because if we treat
dihedral space as Euclidean, we 
can use off-the-shelf techniques and software. 

As an example, we analyze the ground-truth subset of the MPI FAUST dataset. The entire dataset include 300 human 3D laser scans in a wide range of poses~\cite{faust}, and it is
intended as a benchmark for registration methods. 
Its ground truth subset is given as a set of embeddings of a 
single topological mesh, representing 10 subjects each in 10 different poses, labeled by subject and pose. Each mesh has approximately 7,000 vertices.

\begin{figure}[h!]
    \centering
    \subfloat[]{\includegraphics[width=1.0\linewidth]{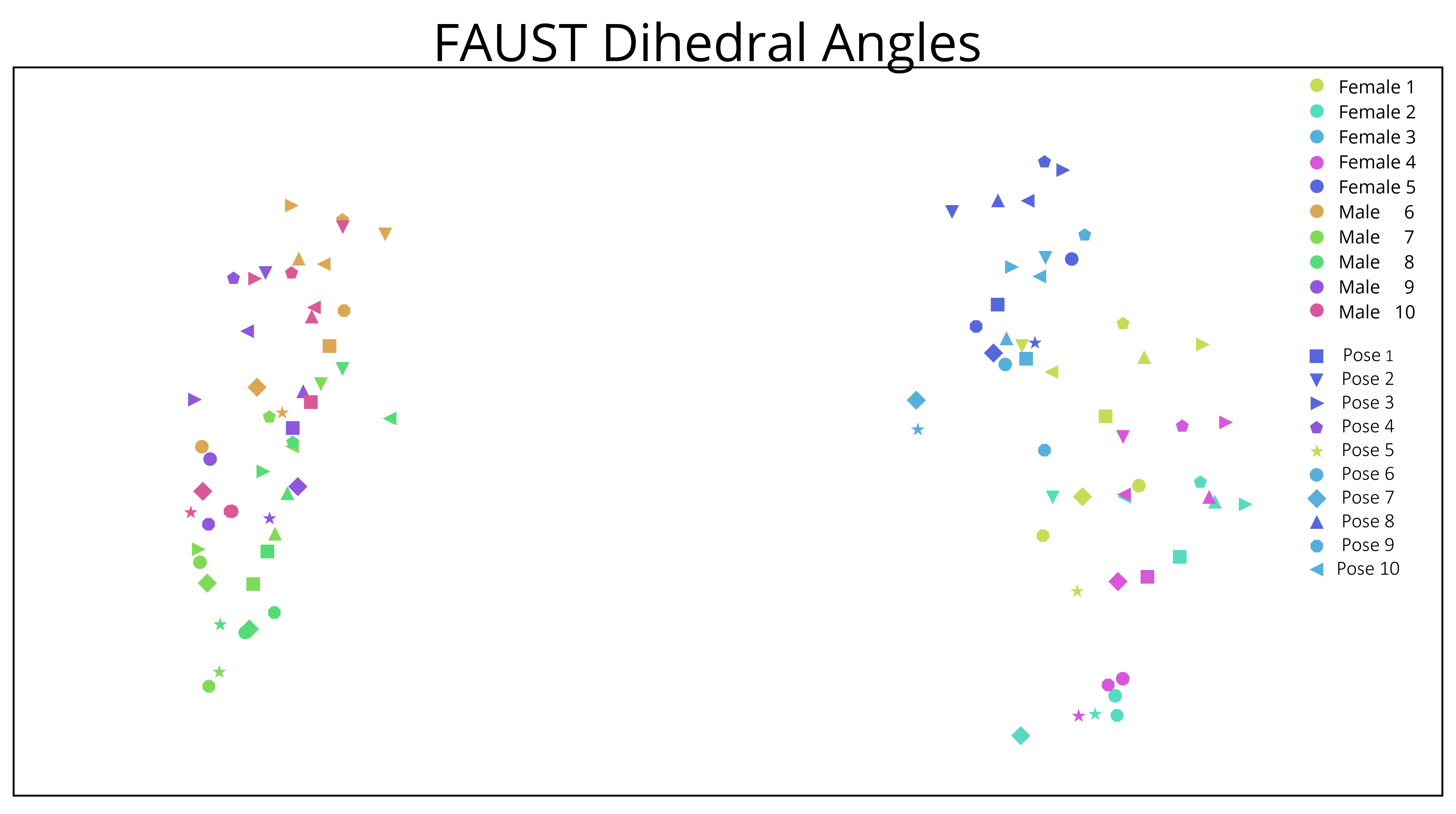} \label{subfig:faust_mf}}\\
    \subfloat[]{\includegraphics[width=1.0\linewidth]{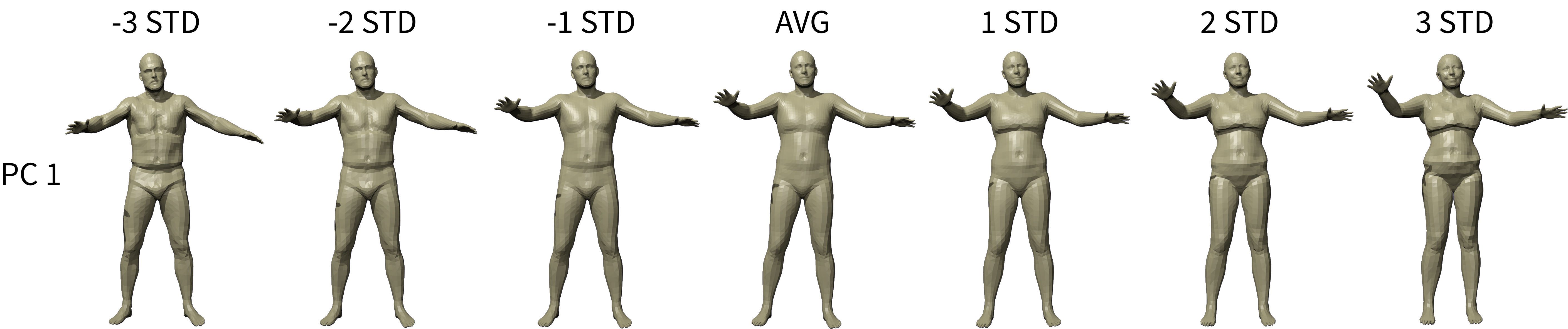} \label{subfig:faust_all}}

    \caption{ The top principal component in dihedral shape space for the FAUST human body shape data describes the fundamental shape difference between male and female bodies.
        When we plot the first two principal components (a) we clearly see the separation between the group of male and female subjects.
        In Figure (b) we warp the average shape in the direction of the first principal component, by adding multiples of it to the average shape.
        All of the figures are reconstructed using the same edge lengths.
        Each edge length is averaged over the whole input set.
    }
    \label{fig:faust}
\end{figure}

In the dihedral space, we used PCA to reduce the dimensions of the ground-truth dataset; a scatterplot on these first two 
principal coordinates is shown at the top in Figure~\ref{fig:faust}.
There are two distinct 460 clusters that correspond to gender, demonstrating the fact that in dihedral space the most salient features are those that reflect body shape rather than pose or size; this is not true,
for example, in the Euclidean 
space formed by the $3n$ vertex coordinates. 
Variation of the body shape along the first principal component in dihedral space is shown at the bottom of Figure~\ref{fig:faust}.
The meshes in this visualization were created using the initial approximate least-squares reconstruction from dihedrals and edge lengths of Appendix~\ref{init}, without optimization, using in every case the average edge lengths over the entire corpus of 100 scans and only varying the input dihedrals
along the principal component.
This shows that quite large changes in shape can be visualized without changing the input edge lengths.

\section{Discussion}

There is a clear gap between the very 
basic level of  our mathematical understanding of the 
dihedral vectors of mesh embeddings and 
the potential reflected in our experimental 
work.  
This suggests several conjectures,
perhaps the most important being,
\begin{conj}
There is at most one set of inner face angles
consistent with an embedding of a mesh
realizing a given vector of dihedral angles.
\end{conj}

\bibliographystyle{amsalpha}
\bibliography{dihedral}

\begin{appendices}

\section{Derivation of the vertex equation}
\label{sec:vertEqn}

First, we need to review some
material on the derivatives of the rigid motions.
To warm up, let us consider translation.
As $t$ goes from zero to one, the coordinate vector $P_0$ changes to $P_0+b$, meaning the vector
$(b_x, b_y, b_z)$ is added to every component of $P_0$.
At time $t$, we have $P(t) = P_0 + tb$.
The derivative then is $dP(t)/dt = b$.

Rotations are more interesting.
Let the matrix $R(\alpha)$ represent the rotation through the axis $(r_x,r_y,r_z)$ (through the origin) by
angle $\alpha$.
At time $t$, we have $P(t) = R(t \alpha) P_0$, that is, the angle of rotation increases with $t$ but the axis stays the same.
So
\begin{eqnarray}
\frac{dP(t)}{dt} & = & \frac{dR(t \alpha)}{dt} P_0 + R(t \alpha) \frac{d P_0}{dt} \\
& = & \frac{dR(t \alpha)}{dt} P_0.
\end{eqnarray}

Interestingly, the derivative
$$
\frac {dR(t \alpha)}{dt} = S R(t \alpha) = (\alpha r_x,\alpha r_y,\alpha r_z) \times R(\alpha t)
$$
is a $3\times3$ matrix, where $S$ is the matrix
$$
\begin{vmatrix}
0 & -\alpha r_z & \alpha r_y \\
 \alpha r_z & 0 & -\alpha r_x  \\
 -\alpha r_y & -\alpha r_x & 0
 \end{vmatrix}
$$
which performs the cross-product.
The vector $\omega = (\alpha r_x,\alpha r_y,\alpha r_z)$ is the axis of rotation of $R$,
is known as the angular velocity vector;
the actual angular velocity of a point $p$
at time $t$ undergoing the rotation, however, is represented by the value of the
derivative at $t$,  $(\omega \times R(\alpha t) ) p_0$.

Notice that $(R_a \omega_b) \times p \neq R_a S_b p$; this is easy to see since the
matrix on the right-hand-side does not
have zero diagonal, and the matrix on the left does.
In fact, the correct transformation is
$(R_a \omega_b) \times p = R_a S_b R_a^T p$,
and we can say $(R_a \omega_b) R_a = R_a S_b$, since $R_a R_a^T = I$.
This property comes in handy when working with the derivative of a series of rotations, as follows.
Say
$$
R_d = R_a R_b R_c
$$
Then
\begin{eqnarray*}
\frac {dR_d}{dt}  & = & \frac {dR_a }{dt} R_b R_c + R_a  \frac {dR_b }{dt} R_c + R_a  R_b \frac {dR_c }{dt} \\
S_d R_d & = & S_a R_a R_b R_c + R_a S_b R_b R_c + R_a  R_b S_c R_c \\
\omega_d \times R_d & = & \omega_a \times R_a R_b R_c + (R_a \omega_b) \times R_a R_b R_c + \\
& & (R_a  R_b \omega_c) \times R_a R_b R_c
\end{eqnarray*}
and hence
$$
\omega_d = \omega_a + (R_a \omega_b) + (R_a  R_b \omega_c)
$$
Notice that the vectors $\omega$ are given in the local coordinate system, so that the multiplications by the preceding rotations
in the equation above are transforming them into the global coordinate system.

\section{Derivation of the cotangent equation at a vertex}
\label{sec:cotVert}

Let $p_i$ be a vertex, and consider the vertices of its one-ring, $p_j, p_{j+1}$, etc.
Using the Law of Sines, we have
$$
\frac {\sin \beta_{i,j,j+1}} {\sin \beta_{i,j+1,j}} = \frac {l_{i,j+1}} {l_{i,j}}
$$
Going around the one-ring,
$$
\prod_j \frac {l_{i,j+1}} {l_{i,j}} = 1
$$
and so
$$
\prod_j \frac {\sin \beta_{i,j,j+1}} {\sin \beta_{i,j+1,j}} = 1
$$
Taking the natural logarithm, we have
$$
\sum_j \ln \sin \beta_{i,j,j+1} - \ln \sin \beta_{i,j+1,j} = 0
$$
Next we take the derivative.  We have $(\ln x)' = 1/x$ and $(\sin x)' = \cos x$, so we get
$$
\sum_j \frac {\cos \beta_{i,j,j+1}} {\sin \beta_{i,j,j+1}} \beta'_{i,j,j+1} - \frac {\cos \beta_{i,j+1,j}}
{\sin \beta_{i,j+1,j}}  \beta'_{i,j+1,j} = 0
$$
or
$$
\sum_j \cot \beta_{i,j,j+1} \beta'_{i,j,j+1} - \cot \beta_{i,j+1,j}  \beta'_{i,j+1,j} = 0
$$

\section{Derivation of Equation~\ref{eqn:cot}}
\label{sec:cot}

We know that
$$
(p_i - p_j) \times (p_i - p_{j+1}) = || p_i - p_j || || p_i - p_{j+1} || \sin \beta_{j,i,j+1} n_{j,i,j+1}
$$
We also know that
$$
(p_i - p_j) \cdot (p_i - p_{j+1}) = || p_i - p_j || || p_i - p_{j+1} || \cos \beta_{j,i,j+1}
$$
So we can write
$$
[ (p_i - p_j) \times (p_i - p_{j+1} ) ]  \cos \beta_{j,i,j+1}  = 
$$
$$
[ (p_i - p_j) \cdot (p_i - p_{j+1} ) ]  \sin \beta_{j,i,j+1} n_{j,i,j+1} 
$$
and hence
$$
[ (p_i - p_j) \times (p_i - p_{j+1}) ]  \cot \beta_{j,i,j+1}  = 
$$
$$
[ (p_i - p_j) \cdot (p_i - p_{j+1})]  n_{j,i,j+1}
$$

\section{Initialization algorithm}
\label{init}

We can initialize the dihedral angle morph in Section~\ref{sec:dihedral_exp} by computing an approximate embedding of the interpolated point.

We approximate the embedding by linearly interpolating the dihedral angles and the edge lengths between the two inputs;
we can construct a mesh that satisfies both in a least-squares sense.
The method we use combines the ideas of \cite{baek2015isometric, kircher2008free, lipman2005linear, winkler2010multi}.
At each vertex $p_i$, we construct a least-squares approximation to its star (the set of triangles containing $p_i$), achieving the desired dihedrals but introducing error in the edge lengths opposite $p$.
We also define an arbitrary canonical coordinate system $F_i$ at each vertex $p_i$.
For every two stars at $p_i$ and $p_j$ connected by an edge $e_{ij}$, we find the three dimensional rotation $R_{ij}$ that takes $F_i$ to $F_j$ when the two stars are merged.
This gives us a relative rotation along each edge.
We use these to solve for a global orientation at each vertex:

$$\min \sum_{e_{ij}}||G_i R_{ij} - G_j||^2_F$$

where $||\cdot||_F$ indicates the Frobrenious norm.
This is a least-squares solve for the $R_{ij}$. 
Because of numerical error and the poor conditioning of the system, we may end up with ``rotations'' $R_{ij}$ that are not actually orthonormal. 
Following~\cite{mao1986optimal}, we correct these using the singular value decomposition. 
This produces a set of global rotations aligning the coordinate frames at every vertex.
Given a good set of $G_{i}$ matrices, we can then use them to reconstruct vertex positions, using

$$\min \sum_{\vec{e}_{ij}} || (p_i - p_j) + G_{i} p_{ij}||^2$$

Here $p_{ij}$ represents the position of the copy of vertex $p_j$ in the original coordinate frame at $p_i$;
as transformed by the global rotation, it should be equal to $p_i - p_j$.
Again, this is a least-squares computation.

\end{appendices}
\end{document}